\documentclass[12pt]{article}
\usepackage{graphicx}
\begin{document}

Social Effects in Simple Computer Model of Ageing
\bigskip

Dietrich Stauffer*$^1$ and Jan P. Radomski$^2$

\bigskip
$^1$ Institute for Theoretical Physics, Cologne University, D-50923 K\"oln, 
Euroland

\bigskip
$^2$ Interdisciplinary Center for Mathematical and Computational Modeling,
Warsaw University, Pawi\'nskiego 5A, Bldg. D, PL-020106 Poland

\bigskip
* Corresponding author stauffer@thp.uni-koeln.de, fax +49 221 470 5159

Running Title: Social Effects in Simple Ageing Model

Keywords: fecundity-survival trade-off, Monte Carlo simulation, population
genetics

\bigskip
Abstract:  
A simple evolutionary model for biological ageing is modified such that
it requires a minimum population for survival, like in human society.
This social effect leads to a transition between extinction and survival
of the species.

\bigskip
Most computer simulations of biological ageing use the Penna model with 
several free parameters [1,2]. Recently Stauffer [3] in a general review
of the Penna model suggested a simpler alternative. (For other applications
of this simpler model see appendix 1). For acceptance in the
medical community [4] it seems important to include specific human aspects,
like the social structures of technological societies. Such an attempt is
made here, similar in spirit but different in details to the simulation of
wolf packs in the Penna model [5].
 
In the simple alternative [3], each living individual every year produces at 
most one offspring which apart from small mutations by $\pm 1$ year has the
same mimimum reproduction age $a_m$ and the same genetic death age $a_d$
as the parent. The time unit is called a ``year'' here, it may correspond four 
years for humans and to days or other time intervals for other species. 
The probability to have one offspring is
assumed to be $(1 + 0.08)/(a_d(i)-a_m(i) + 0.08)$, i.e. the birth rate is the
smaller the longer the reproductive life of the parent is: fecundity-survival
trade-off [6]. Individuals die
if their genetic death age $a_d$, different for different species, is reached;
they also die every year with a Verhulst probability $V(N) = N/N_{max}$ due to
food and space restrictions, where $N$ is the actual population at that time, 
and $N_{max}$ an input parameter, often called the carrying capacity (see  
appendix 2). 

(No individual is allowed to live beyond 32 ``years'' and no mutations are
allowed to make $a_d > 32$ or $a_d \le a_m$; similarly we require $0 \le a_m < 
32$.)

Now we modify this model to include social effects: humans need other humans
to survive. Thus not only a too high population is bad (as simulated
by the above Verhulst factor), but also a too low population. We thus take
$$ V(N) = N_{min}/N + N/N_{max} \quad (N_{min} < N < N_{max}) $$
where $N_{min} (\ll N_{max})$ is a new parameter which is zero in the old model 
[3]. Fig.1 shows this probability for the parameters of our simulations, and
Fig.2 presents some simulations. The actual population is not in the minimum
of $V(N)$. If $N_{min}$ is not small enough compared with $N_{max}$, the 
population dies out. A fortran program is available from 
stauffer@thp.uni-koeln.de, a C version from janr@icm.edu.pl.
Appendix 3 discusses some parameters.
\begin{figure}[hbt]
\begin{center}
\includegraphics[angle=-90,scale=0.5]{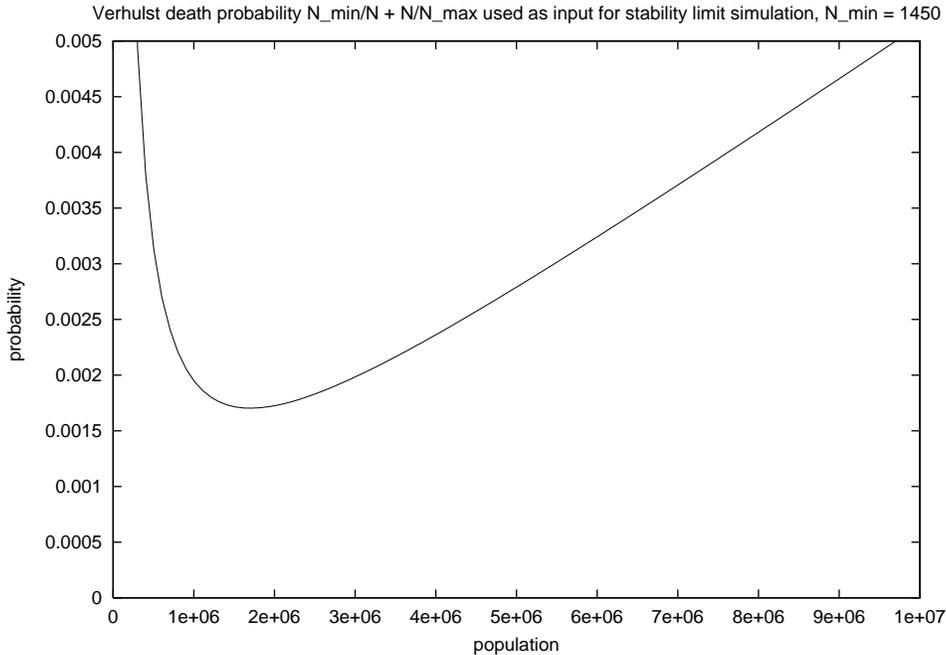}
\end{center}
\caption{
Example of the Verhulst dying probability, making both a too small and 
a too large population unfavourable.
}
\end{figure}    

\begin{figure}[hbt]
\begin{center}
\includegraphics[angle=-90,scale=0.5]{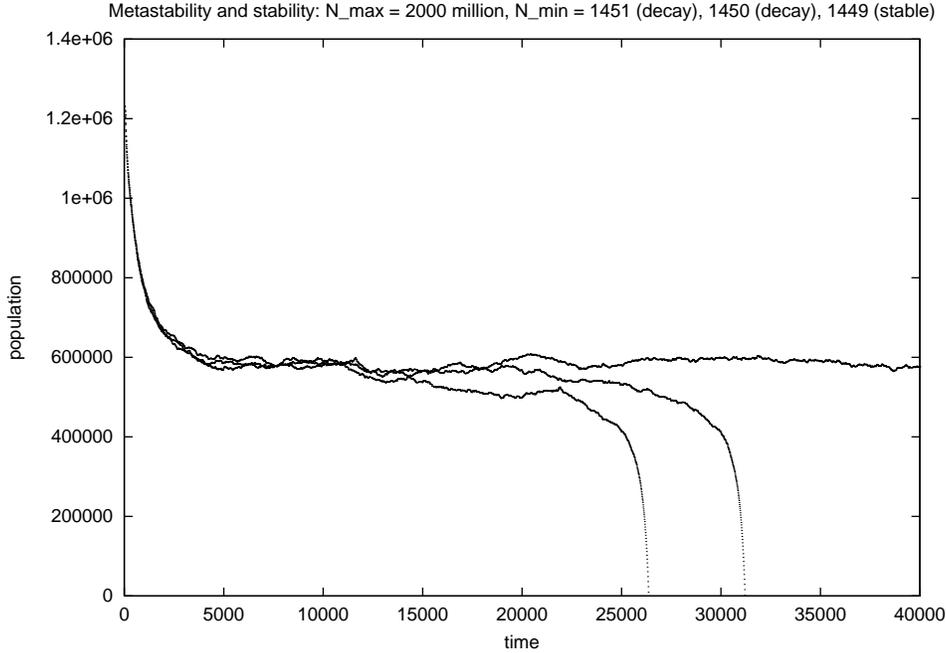}
\end{center}
\caption{
Transition between stable and decaying populations; $N_{min}$ varies
by about a tenth of a percent between these curves.
}
\end{figure}    

\begin{figure}[hbt]
\begin{center}
\includegraphics[angle=-90,scale=0.5]{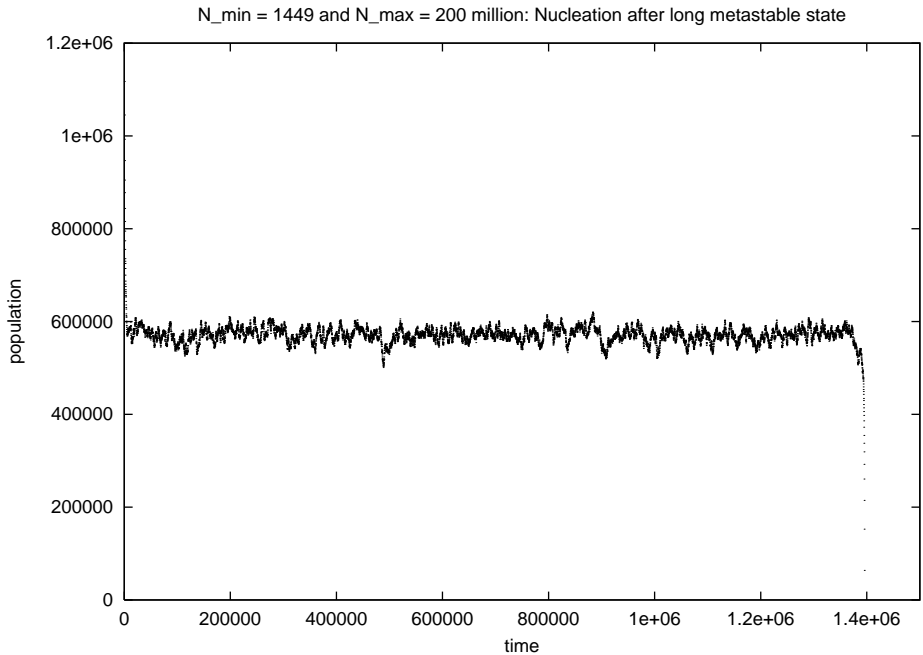}
\end{center}
\caption{
Metastable plateau decays after 1.4 million years
(continuation of the flat curve of Fig.2).
}
\end{figure}    

The transition between a surviving and a dying population is quite sharp, Fig.2.
If we wait long enough then even a population which first seemed stable can
become unstable: Fig.3 shows the rather flat curve of Fig.2 for 35 times more
years, when quite suddenly it decays. This sudden decay comes from the negative
feedback of the social term  $N_{min}/N$ in our Verhulst probability: If $N$ 
happens to have a strong downward fluctuation, this dying probability 
$N_{min}/N$ becomes larger, thus further decreasing the population. 
The mortality function versus age, and the distributions of $a_m$ and $a_d$
are not affected by our age-independent Verhulst factor, and thus the same
as in Ref. 3.
 
(Introduction of a maximum reproduction age can easily kill the population;
setting it to 30 when the maximum possible age is 32
reduces the average age of genetic death from 18.6 to 18.4.)

Similar effects are seen for the traditional Penna model [1,2], 
modified by the above Verhulst factor [7]. 

In the following appendices we give more details on some of the above parameters
and assumptions, and summarize other applications of this model. 
\bigskip 

We thank T. Klotz for suggesting this work, and S. Moss de Oliveira for 
communicating her results [7]. 
JPR was partially supported by the KBN-115/E343/2001/ICM grant. 
\vskip 1cm

\centerline {Appendix 1: Other applications}
Before medical conclusions are drawn from these simulations, more experience
needs to be gained with this model in other aspects. Thus we shortly summarize
from the literature other tests of the simple ageing model of ref.[3].

The catastrophic senescence of Pacific Salmon, that is the fish dying soon after
producing offspring, is one of the crucial facts of nature which a theory
has to explain. It comes out reasonably [8] in the simple ageing model, if one
assumes that the fish have offspring only once in their lifes, and all at the
same age. However, the transition between survival before reproduction, and
death after reproduction was not as sharp in [8] as in the earlier simulations
with the more complicated Penna model [9].

In 1993 the Northern Cod vanished off the coast of Newfoundland (Grand Banks),
and the Penna model [10] could explain this effect through overfishing. Fig.4
shows that also the simpler model [3] can explain extinction through this 
overfishing: First we let the fish reach their stationary age distribution
without any human fishing. Then, after 10000 iterations, humans take away 
one percent of the fish population, resulting in a drop of the surviving 
population to a lower but still stable value (center part of Fig.4). Finally,
after 20000 iterations, another one percent are fished away, and now the 
population shrinks exponentially to zero in the right-most part of this figure.
 
\begin{figure}[hbt]
\begin{center}
\includegraphics[angle=-90,scale=0.5]{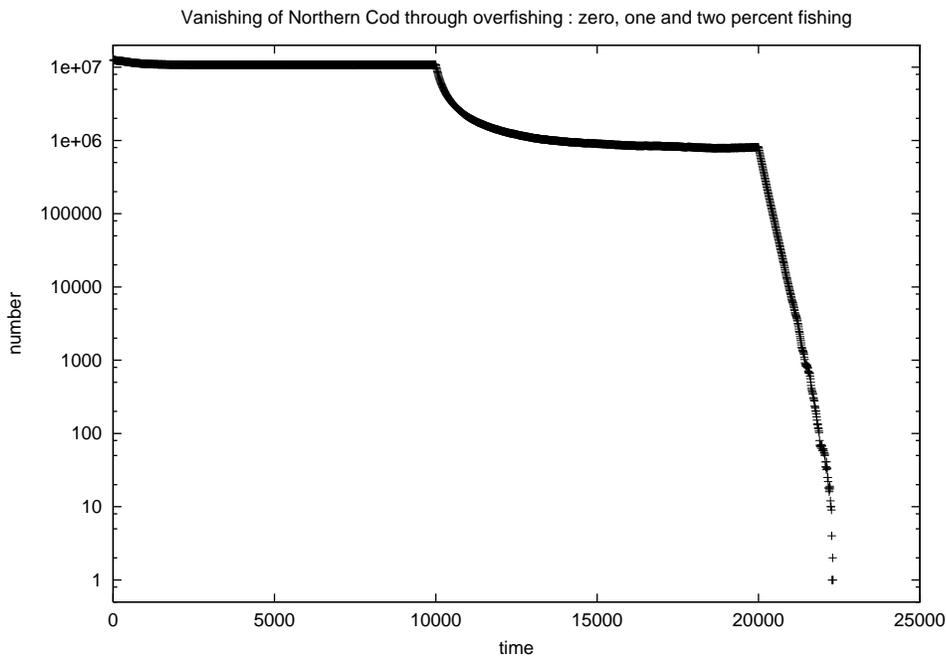}
\end{center}
\caption{
Overfishing in simple model, without social effects. First part: no 
fishing; second part: fishing of one percent gives lower but still stable
population; third part: fishing increased to two percent leads population
to extinction.
}
\end{figure}    

The Penna model was put onto a lattice [11] to simulate e.g. the effect that a 
tree cannot grow on the site where another tree already grows. Similar results
were obtained [12] if the simple model [3] is put onto a lattice. This lattice
model allows to avoid an explicit Verhulst factor (see next appendix), as 
pointed out by Makowiec [11].
  
The mortality not only of humans increases roughly exponentially with age
(Gompertz law) if childhood diseases are overcome. This increase was found
nicely in the Penna model [1,2] but is followed less accurately in the
simple model; the latter results follow [12] the Gompertz law better
if death does not occur abruptly 
for all individuals at their genetic death age, but with some probability
increasing in a narrow age interval around the genetic death age [13].
Alternatively, reasonable agreement with the Gompertz law is obtained by 
enhancing the birth rate for old individuals [14].

Generalization to sexual reproduction also gave reasonable results similar
to the esexual case [12]. In summary, 
inspite of its simplicity the model of ref.3 passed already the most crucial 
tests, which have been before applied successfully to the more complicated Penna
ageing model. However, some of the results are not as good as for the Penna
model.

\bigskip
\centerline{Appendix 2: Verhulst Factor}

The simple asumption that each individual gives rise to $b$ births in each
small time interval, and dies with a constant probability $d$ during each time
interval, leads to a differential equation $dN(t)/dt = aN(t)$ or discrete
equation $N(t+1) - N(t) = a N(t)$ for the number $N$ of individuals in 
a population, with $a=b-d$. This differential equation is solved by an
exponential function, $N(t) = N(0) \exp(at)$, leading to extinction ($ a < 0$)
or unlimited growth ($a > 0$). The latter is unrealistic, and thus since the 
19th century the equation is improved to $dN(t)/dt = aN(t) V(t)$
where $V(t) = 1-N/N_{max}$ is the Verhulst factor, taking into account 
limitations in food and space. The carrying capacity $N_{max}$ is in this
simple case the infinite-time limit of the population: $N(t \rightarrow \infty)
\rightarrow N_{max}$. In more complicated models, like the present one of the 
Penna ageing model, this Verhulst factor $V = 1-N/N_{max}$ is only one of 
several elements, and the infinite-time limit of the population no longer
has to agree with $N_{max}$; nevertheless $V$ is a computationally useful and
biologically realistic ingredient to take into account the limits of growth.

Special ``chaos'' effects may occur [15] in the discrete approach 
$N(t+1) - N(t) = a N(t)V(t)$ (logistic equation) if the
population changes drastically from one time step to the other. In our 
simulations the populations varied smoothly with time and obeyed $N \ll N_{max}$
thus avoiding these complications. They could appear at higher birth rates,
while we assumed one birth attempt per time step for each suitable individual.

In most applications as well as in the present paper, this Verhulst factor 
applies equally to all ages. It may be more realistic biologically to apply it
only to babies, in which case it acts like a reduction of the birth rate. In
the Penna ageing model, the restriction of the Verhulst factor to babies
did not give major changes in the equilibrium results [16]; it sometimes gives
an unrealistic time dependence in how the equilibrium population is approached
and thus was not applied in the present model. 

\bigskip
\centerline {Appendix 3: Parameters}

The main advantage of this simple ageing model compared to the more realistic 
Penna model is that it has few free parameters. We set the birth rate to 
one in the sense that each mature individual in each iteration tries to have
one offspring, with the probability given in the main text and depending on
the two characteristic ages. Then the population survives but is close to 
extinction, as is perhaps realistic; thus the population $N$ is far below
$N_{max}$. With a birth rate (in the above sense) of two, the population was
higher and no longer close to extinction, but then the increase of mortality 
with increasing age turned out to be less realistic.

Since the model is close to extinction, the ratio $N_{min}/N_{max}$ must be 
taken as very small; in order of magnitude the actual population $N$ is the
geometric mean of $N_{min}$ and $N_{max}$, where the Verhulst death probability 
$N_{min}/N + N/N_{max}$ has its minimum. We took it such that both survival and 
extinction could be observed by slight changes in $N_{min}$.

The free parameter 0.08 in the birth probability was introduced originally to
avoid a wrong divergence if $a_d = a_m$ in the birth probability 
$1/(a_d-a_m)$. It was then taken such that the population is close to 
extinction and that the age distribution does not extend beyond the maximum
age of 32 time intervals, traditional for many Penna model studies. This
parameter 0.08 from the simple ageing model without the social effects discussed
here was then taken over without any change. So the new effects seen here
are due to our introduction of $N_{min}$ and not due to this parameter 0.08.
When we omit this parameter 0.08 and omit also the restriction of a maximum
age of 32, we still get a stable population and reasonable results in the 
simple model, also with the present social effects. 

\parindent 0pt
\bigskip
[1]  T.J.P. Penna, J. Stat. Phys. 78, 1629 (1995).

[2] S. Moss de Oliveira, P.M.C. de Oliveira, D. Stauffer, {\it Evolution, 
Money, War and Computers}, Teubner, Stuttgart and Leipzig 1999.

[3] D. Stauffer, {\it Biological Evolution and Statistical Physics} (Dresden, 
May 2000), edited by M. L\"assig and A. Valleriani, Springer, Heidelberg 
and Berlin 2002.

[4] T. Klotz, private communication.

[5] S.J. Feingold, Physica A 231, 499 (1996) and Int. J. Mod. Phys. C 9, 295
(1998); S. Cebrat, J. K{\c a}kol, Int. J. Mod. Phys. C 8, 417 (1997); D.
Makowiec, Physica A 245, 99 (1997). For simple model see T. Shimada, Int. J. 
Mod. Phys. C 12, No. 7 (2001). For inbreeding effects see
A.O. Sousa, S.M. de Oliveira, A.T. Bernardes, Physica A 278, 563 (2000).

[6]  C.K. Galambor, T.E. Martin. Science 292, 494 (2001)

[7] S. Moss de Oliveira, private communication, April 2001.

[8] H. Meyer-Ortmanns, Int. J. Mod. Phys. C 12, 319 (2001).

[9] T.~J.~P.~Penna, S.~Moss de Oliveira and D.~Stauffer, Phys.~Rev.
~E 52, R3309 (1995).

[10] S. Moss de Oliveira, T.J.P. Penna, D. Stauffer: Physica A 215, 298 (1995).
 
[11] A.O. Sousa, S. Moss de Oliveira, Eur. Phys. J. B 9, 365 (1999);
D. Makowiec, Physica A 289, 208 (2000); see also M.S. Magdo\'n, A.Z.
Maksymowicz, Physica A 273, 182 (1999).

[12] A.O. Sousa, S. Moss de Oliveira, and D. Stauffer, Int. J. Mod.  Phys. C 
(submitted).

[13] J. Thoms, P. Donahue, N. Jan, J. Physique I 5, 935 (1995).

[14] D. Makowiec, D. Stauffer and M.  Zieli\'nski, Int. J. Mod. Phys. C 12,
No. 7 (2001).

[15] A. Raab, J. Stat. Phys. 91, 1055 (1998); A.T. Bernardes, J.G. Moreira
and A. Castro-e-Silva, Eur. Phys. J. B 1, 393 (1998).
 
[16] J.S. S\'a Martins and S. Cebrat: Theory in Biosciences 119, 156 (2000); see
also J. D{\c a}bowski, M. Groth, D. Makowiec: Acta Phys. Pol. B 31, 1027 (2000).

\end{document}